# Recognition and Ranking Critical Success Factors of Business Intelligence in Hospitals - Case Study: Hasheminejad Hospital

Marjan Naderinejad and Mohammad Jafar Tarokh and Alireza Poorebrahimi

Department of Information Technology Management, Science and Research branch,
Islamic Azad University, Tehran, Iran

**ABSTRACT**

*Background and Aim: Business Intelligence, not as a tool of a product but as a new approach is propounded in organizations to make tough decisions in business as shortly as possible. Hospital managers often need business intelligence in their fiscal, operational, and clinical reports and indices. Recognition of critical success factors (CSF) is necessary for each organization or project. Yet, there is not a valid set of SCF for implementing business intelligence. The main goal of recognition and ranking CSF is implementation of a business intelligent system in hospitals to increase success factor of application of business intelligence in health and treatment sector.*

*Materials and Methods: This paper is an application and descriptive-analytical one, in which we use questionnaires to gather data and we used SPSS and LISREL to analyze them. Its statistical society is managers and personnel of Hasheminejad hospital and case studies are selected by Cochran formula.*
*Results: The findings show that all three organizational, process, and technological factors equally affect implementation of business intelligence based on Yeoh & Koronis approach, where the assumptions are based upon it. The proposed model for CSFs of business intelligence in hospitals include: declaring perspective, goals and strategies, development of human and financial resources, clarification of organizational culture, documentation and process mature, management support, etc.*

*Conclusion: Business intelligence implementation is affected by different components. Center of Hasheminejad hospital BI system as a leader in providing quality health care, partially succeeded to take advantage of the benefits the organization in passing the information revolution but the development of this system to achieve intelligent hospital and its certainty is a high priority, thus it can`t be said that the hospital-wide BI system is quite favorable. In this regard, it can be concluded that Hasheminejad hospital requires practical model for business intelligence systems development.*

## 1. INTRODUCTION

IT includes a part of daily life for most of people. Today, business intelligence is one of the modern mechanisms to increase competitive advantage and overcome competitors. Health and treatment sector is one of the challengeable and sensible areas for intelligence [1].

In fact, BI is defined as application of artificial intelligence in administration of hospitals and medical affairs, including diagnosis and treatment. For example, since a hospital manager deals with a lot of data, he is not able to analyze it, and since there is not enough opportunity, here BI is applied. BI receives this data, processes it, and results a hospital dashboard for the manager. If a specialized physician has less concerns, he can assign more time to diagnose diseases, so treatment process become better [2].

Certainly, applying BI brings many advantages for treatment services. If a physician assigns more time to each patient, then there is lower error probability and more cure success [2].

Treatment Services Providers (TSP) use BI solutions to analyze clinical data, to measure performance, and to report, but it is not enough having good software. BI has no benefit without data. The following points are very important in BI:

1. Data quality
2. Identification of repeated data
3. Integration of data for all available sources
4. On-time reporting

Availability of information to access clinical, financial and administrative data rapidly, helps managers for decision-making. Organizations shall accept that their life philosophies have changed and survival will not merely mean continuous profit. They should seek competition because today little companies act traditionally. If an organization wants to proceeds others, it must be familiar with new game rules. Different organizations with different business encounter different problems for applying current data in sale, stock, and financial systems optimally. [3]

Nonetheless, organizations must provide necessary capacities to use current data and take important steps and develop IT in their organizations.

It was frequently observed that an organization commits a large project, but it will not benefit it after a large amount of costs. So it is better to perform a sample project with a smaller dataset and to use key factors. [4]

Therefore, relationship and effect of CSF in implementation of BI in health and treatment sector, especially in hospitals, became one of the concerns of the researcher. Thus, the researcher intends to identify a framework of CSFs to implement business intelligence considering their priority needs BI implementation raises the following questions:

Q1: What are the organizational factors for implementation of BI in hospitals?
Q2: What are the process factors for implementation of BI in hospitals?
Q3: What are the technological factors for implementation of BI in hospitals?

This research was carried out to find answers to the above questions and to identify a framework of CSFs to implement business intelligence successfully. The rest of this paper is organized as follows: Section 2 consists of a literature review of CSF and BI definitions and list of CSFs. Section 3 discusses the research objectives and methodology. Section 4 describes the questionnaire results and an analysis including hypothesis testing of the main research question of factor extraction and ranking. Finally, concludes the research work and its main results and limitations, and proposes directions for future research.

## 2. LITERATURE REVIEW

### 2.1. Concepts of CSFs

There are many features and conditions, or variables that significantly affect success of an organization if they are managed precisely [5]. CSFs are used to identify and prioritize business needs and technical systems [6]. These factors help improvement of processes. They will be more effective if they are used by their importance in each implementation step [7]. CSFs had been defined in different areas to obtain success [8].

Also, CSFs are used to anticipate future successes of an organization. CSFs are very useful in organizational managerial studies, and these areas have not been studied in our country yet.

## 2.2. Concept of BI

BI includes a broad range of applications and technologies to gather data and knowledge to generate queries for analyzing and organization to make precise and intelligent decisions [9].

The concept of BI was suggested following disorders of MISs. MISs has grown theoretically, but they never could respond needs of organizations. BI is a collection of abilities, technologies, tools, and strategies that help managers to better understand their organizations. BI tools provide views from past, now, and future conditions. Implementing BI strategies diminishes the gap between middle managers and top managers by communication and necessary information will be provided for them immediately in a qualitative manner. Also, experts and analysts can improve their activities by simple facilities and access better results [3,12].

Mister (2009) suggests challenges of successful implementation of BI as: identification of key needs, data source quality evaluation, ensuring flexibility in final implementation; and he remarks steps of a successful implementation of BI as: large thinking and small beginning, using frequent techniques in definition of needs, and work domain [10].

## 2.3. Critical Success Factors for BI

Ariyachandra and Watson (2006), analyzing CSFs for BI implementation, take into account two key dimensions: process performance (i.e., how well the process of a BI system implementation went), and infrastructure performance (i.e., the quality of the system and the standard of output). Process performance can be assessed in terms of time-schedule and budgetary considerations. Whereas infrastructure performance is connected with the quality of system and information as well as this system use [11,12].

According to Yeoh and Koronios (2010), CSFs can be broadly classified into three dimensions: organization, process, and technology. Organizational dimension includes such elements as committed management support and sponsorship, a clear vision, and a well-established business case. In turn, the process dimension includes business-centric championship and balanced team composition, business-driven and interactive development approach and user-oriented change management. Technological dimension regards such elements as business-driven, scalable and flexible technical framework, and sustainable data quality and integrity.

Their findings show that there are more non-technical indices that are more effective than technical ones on BI systems [12,13].

Following Yeoh & Koronis, in this research, we study CSFs in three areas.

## 3. RESEARCH OBJECTIVES AND METHODOLOGY

To answer the research questions posed in Section 1, several research objectives were determined. The main objective was to determine and rank CSFs to implement BI in hospitals. Subordinate research objectives were to recognition and ranking of organizational, process and technological factors affecting implementation of BI in hospitals.

This study is conducted in 2 phases:

1) Literature review: Articles and documents gathered in the previous researches and investigating factors affecting on BI implementation and preparing questionnaire to recognition CSF of BI implementation.
2) Field study: distributing questionnaire, collecting required data and analyzing and aggregating the results framework are based on ''factor analysis'', and concentrate on the extraction and identification of the

After distributing questionnaires, collected data were entered into SPSS and LISREL. Finally, the most important factors and their effect were made clear through ranking.

## 3.1. Design of the questionnaire

A questionnaire was designed and structured in two sections. Information related to the basic profile was requested at the beginning of the questionnaire. In the second part, 56 questions were asked to measure their attitudes, based on the importance of the CSF of BI implementation. The selected responses were evaluated on a ''Likert Scale'' and the responses could be: Vital importance, Very Important, Important, Low-importance, Devoid of significance, Negligible. In other words, the second part of the questionnaire measures their opinions about the importance of each CSF of implementation.

## 3.2. Methodology

Method of sample selection in this study is by using Cochran's formula. No. of sample size is 424 persons, 201 persons were selected. Selection process method is as follows:

$$z=1.96, \quad p=q=0.5, \quad d=0.1 \quad \Rightarrow \quad n=\frac{\frac{z^2 pq}{d^2}}{1+\frac{1}{N}(\frac{z^2 pq}{d^2}-1)}=\frac{\frac{(1.96)^2 \times 0.5 \times 0.5}{(0.1)^2}}{1+\frac{1}{424}(\frac{(1.96)^2 \times 0.5 \times 0.5}{(0.1)^2}-1)}=201$$

## 4. RESULTS

### 4.1. Hypothesis test

We used average statistical assumption test or One-sample t-test to examine the assumptions. In fact, this test examines the difference between this sample and and hypothesized value. Null hypothesis in all variables, by Likert scale is:

H0:    $\mu = 4$
H1:    $\mu \neq 4$

Table 1. Study of variables

| Variable | Sig. | t | Average | Acceptance |
|---|---|---|---|---|
| Organizational factor | 0.000 | 93.670 | 5.7883 | Confirmed |
| Process factor | 0.000 | 80.009 | 5.7536 | Confirmed |
| Technological factor | 0.000 | 83.537 | 5.7631 | Confirmed |

International Journal of Computer Science & Information Technology (IJCSIT) Vol 6, No 2, April 2014

The results show that assumption 1 is confirmed. All organizational, process, and technological factors affect implementation of Bi in hospitals. For assumption 2, since all factors have the same average, thus sum of two of them is more than the other. Then, the assumption 2 is confirmed; namely, process and organizational factors affect more on implementation of BI in hospitals.

**4.2. CSFs of implementation of BI in hospitals**

According to Friedman Test (table 2), we can say that the average importance levels of organizational, process, and technological factors have not a significant difference in implementation of BI (P>0.05).

Table 2. Results of Friedman Test

| Sub-main variables |
|---|
| Sig. level: 0.467 |

Table 3. Ranking variables

| Variable | Rank average |
|---|---|
| Organizational factor | 2.03 |
| Process factor | 1.94 |
| Technological factor | 2.03 |

As you see in table 3, the importance levels of organizational, process, and technological factors in implementation of BI are the same. Therefore, no factors have priority than the others.

**4.3. CSFs of BI In hospitals by factor**

• Ranking organizational factor components

According to Friedman Test (table 4), we can say that the average importance levels of components of organizational factors have a significant difference in implementation of BI (P<0.05).

Table 4. Results of Friedman Test

| Organizational components |
|---|
| Sig. level: 0.000 |

As you see in table 5, components of perspective, goals, and strategy are more important in implementation of BI, because they have more average rank, but component of management support is less important because it has a lower average rank.

Table 5. Ranking components of organizational factor

| Component | Rank average | Priority |
|---|---|---|
| Perspective, goals, and strategy | 4.75 | 1 |
| Financial resources | 4.13 | 2 |
| Human resources | 4.09 | 3 |
| Organization culture | 3.95 | 4 |
| Leadership | 3.90 | 5 |
| Coincidence of business and IT | 3.65 | 6 |
| Management support | 3.54 | 7 |

International Journal of Computer Science & Information Technology (IJCSIT) Vol 6, No 2, April 2014

• Ranking components of process factor

According to Friedman Test (table 6), we can say that the average importance levels of components of process factor have a significant difference in implementation of BI (P<0.05).

Table 6. Results of Friedman Test

| Process components |
| --- |
| Sig. level: 0.000 |

As you see in table 7, component of process maturity has more importance in implementation of BI, because it has more average rank, but component of project team combination is less important because it has a lower average rank.

Table 7. Ranking components of process factor

| Component | Rank average | Priority |
| --- | --- | --- |
| Process maturity | 3.68 | 1 |
| Methodology | 3.66 | 2 |
| Change management | 3.64 | 3 |
| Frequent development model | 3.47 | 4 |
| Process documentation | 3.36 | 5 |
| Project team combination | 3.19 | 6 |

• Ranking components of technological factor

According to Friedman Test (table 8), we can say that the components of technological factor are:

Table 8. Ranking components of organizational factor

| Component | Rank average | Priority |
| --- | --- | --- |
| Technology and knowledge transfer speed | 3.15 | 1 |
| Data quality | 3.12 | 2 |
| Suitable infrastructure and technology | 2.96 | 3 |
| Application capability | 2.93 | 4 |
| Training and support | 2.84 | 5 |

According to the above table, the component of technology and knowledge transfer speed and the component of data quality have near averages and are more important in implementation of BI, because their average ranks are higher. Also, the component Training and Support is less important, because its average rank is lower.

• Ranking components of BI success

Regarding to the results, the set of CSFs are as table 9. It should be mentioned that some of factors are in the same rank because their average values are near.

Table 9. Ranking components

| Priority | Component |
|---|---|
| 1 | Perspective, goals, and strategy |
| 2 | Financial resources |
|   | Human resources |
| 3 | Organization culture |
| 4 | Leadership |
| 5 | Process maturity |
|   | Methodology |
| 6 | Coincidence of Business and IT |
|   | Change management |
| 7 | Management support |
| 8 | Frequent development model |
| 9 | Process documentation |
| 10 | Project team combination |
|    | Technology and knowledge transfer speed |
|    | Data quality |
| 11 | Suitable infrastructure and technology |
|    | Application capability |
| 12 | Training and support |

## 5. CONCLUSION

Regarding to the obtained results from field studies, analysis of inferential statistics, and study of cause and effect patterns, this model can be implemented in other hospitals, too.

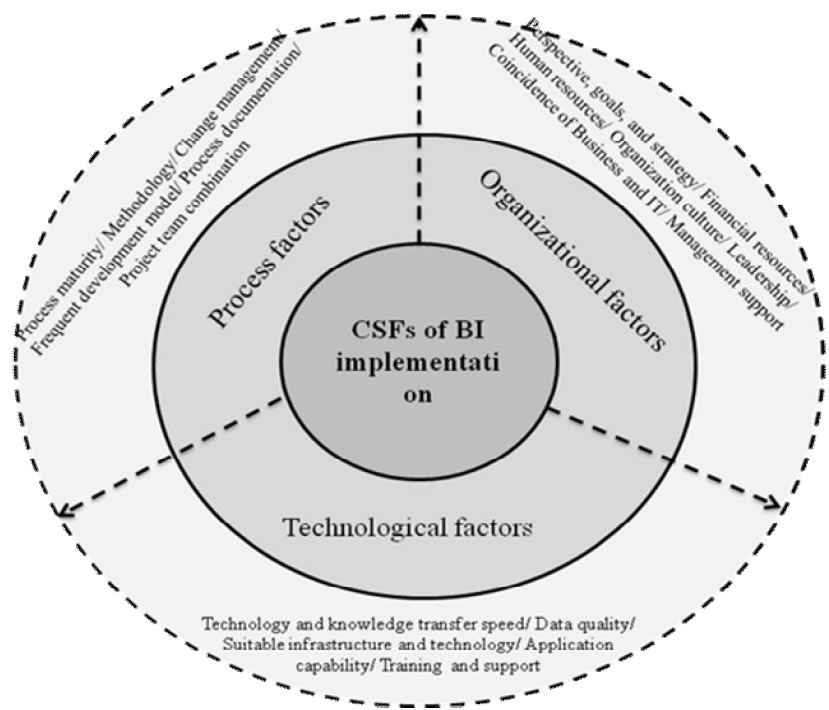

Figure 1. Integrated model of factors and components of CSFs of BI in hospitals

The most important result of this paper is that implementation of BI is affected by different factors. This paper confirms the effects of organizational, process, and technological factors by Yoeh and Coronis Approach. This shows that BI is a discussion in technology and human and social sciences, so rather than IT, management should be viewed as an art, not a mere science.
In the IT era, transformation of administrative organizations of Medical Sciences University and medical centers from their traditional situation into a modern view is very important. Regarding to the results of this research, BI system of Hasheminejad Hospital, which offers qualitative medical services, can utilize its organizational benefits. However, developing this system is very important to attain an intelligent hospital. Thus, we cannot say the BI system in this hospital is in a desirable level.

On the other hand, it must be mentioned that different variables that were categorized in the frame of organizational, process, and technological factors, have different effects. Therefor, Hasheminejad Hospital needs more applicatory patterns to develop its BI system. Study of world hospitals and comparing it's with Iranian hospitals certainly can offer more applicable strategies toward development of BI system in Hasheminejad Hospital and other state hospitals.

The authors believe that this research will enable hospitals to make better decisions for designing, selecting, evaluating and buying BI systems, using criteria that help them to create a better decision-support environment in their work systems. The main limitations of this research include differences between the functionalities of enterprise systems and the novelty of BI in business and health sector. Of course, further research is needed. One important topic for the future is the design of expert systems (tools) to compare vendor products. Another is application of the criteria and factors that we have identified and defined in a framework, in order to select and rank BI systems based on specifications. The complex relationship between these factors and the satisfaction of managers with the decision making process should also be addressed in future research.

## Authors

**Marjan Naderinejad** is postgraduate in Information Technology Management at Science and Research branch Islamic Azad University, Tehran, Iran. Her research interests are business intelligence, information technology, and decision making. She is working in area of IT at Tehran University of Medical and Sciences (TUMS).

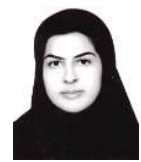

**Mohammad Jafar Tarokh** is the Associate Professor of the Industrial Engineering faculty and head of Postgraduate Information Technology Engineering Group at K. N. Toosi University of Technology, Tehran, Iran. He has been the author and co-author of more than 40 academic papers. His major research interests are: SCM, CRM, e-Commerce, and m-Commerce.

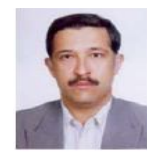

**Alireza Poorebrahimi** is a Ph.D. degree candidate of Industrial Management at Islamic Azad University, Tehran, Iran. He received his MA degree in Industrial Management, from Islamic Azad University, Tehran, Iran. His research interests are Database design, business intelligence, information technology, and information management. He has been the author and co-author of more than 30 papers published in different conferences and journals.

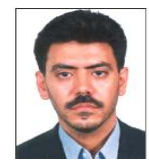